\documentclass[journal]{IEEEtran}
\usepackage{cite}

\usepackage{color,soul}
\usepackage{gensymb}
\usepackage{dsfont}

\ifCLASSINFOpdf
  \usepackage[pdftex]{graphicx}
  \graphicspath{{../pdf/}{../jpeg/}}
  \DeclareGraphicsExtensions{.pdf,.jpeg,.png}
\else
  \usepackage[dvips]{graphicx}
  \graphicspath{{../eps/}}
  \DeclareGraphicsExtensions{.eps}
\fi
\usepackage[cmex10]{amsmath}
\usepackage{amssymb}
\DeclareMathOperator*{\argmaxA}{arg\,max}
\DeclareMathOperator*{\argminA}{arg\,min}
\usepackage{euscript}
\newcommand{\norm}[1]{\left\lVert#1\right\rVert}
\usepackage{multirow}
\usepackage{algorithm}
\usepackage[noend]{algpseudocode}
\usepackage{diagbox}
\usepackage{physics}
\usepackage{tikz}

\ifCLASSOPTIONcompsoc
  \usepackage[caption=false,font=normalsize,labelfont=sf,textfont=sf]{subfig}
\else
  \usepackage[caption=false,font=footnotesize]{subfig}
\fi
\usepackage{url}

\begin{document}
%
\title{A Game-Theoretic Data-Driven Approach for Pseudo-Measurement Generation in Distribution System State Estimation}
%
%
%

\author{
         Kaveh Dehghanpour,~\IEEEmembership{Member,~IEEE,}
         Yuxuan Yuan,~\IEEEmembership{Student Member,~IEEE,}
         Zhaoyu Wang,~\IEEEmembership{Member,~IEEE,}
         Fankun Bu,~\IEEEmembership{Student Member,~IEEE}
\thanks{This work is supported by the U.S. Department of Energy Office of Electricity Delivery and Energy Reliability under DE-OE0000875.

K. Dehghanpour, Y. Yuan, Z. Wang, and F. Bu are with the Department of
Electrical and Computer Engineering, Iowa State University, Ames,
IA 50011 USA (e-mail: kavehd@iastate.edu; wzy@iastate.edu).
}
}
%
%

\markboth{Submitted to IEEE for possible publication. Copyright may be transferred without notice}%
{Shell \MakeLowercase{\textit{et al.}}: Bare Demo of IEEEtran.cls for Journals}
%



\maketitle

\begin{abstract}
In this paper, we present an efficient computational framework with the purpose of generating weighted pseudo-measurements to improve the quality of Distribution System State Estimation (DSSE) and provide observability with Advanced Metering Infrastructure (AMI) against unobservable customers and missing data. The proposed technique is based on a game-theoretic expansion of Relevance Vector Machines (RVM). This platform is able to estimate the customer power consumption data and quantify its uncertainty while reducing the prohibitive computational burden of model training for large AMI datasets. To achieve this objective, the large training set is decomposed and distributed among multiple parallel learning entities. The resulting estimations from the parallel RVMs are then combined using a game-theoretic model based on the idea of repeated games with vector payoff. It is observed that through this approach and by exploiting the seasonal changes in customers' behavior the accuracy of pseudo-measurements can be considerably improved, while introducing robustness against bad training data samples. The proposed pseudo-measurement generation model is integrated into a DSSE using a closed-loop information system, which takes advantage of a Branch Current State Estimator (BCSE) data to further improve the performance of the designed machine learning framework. This method has been tested on a practical distribution feeder model with smart meter data for verification.    
\end{abstract}


%
\IEEEpeerreviewmaketitle

\section{Introduction}
Electric distribution systems have been undergoing radical changes in control and management. The driving force behind these changes can be attributed to higher penetration of distributed renewable resources and employment of Advanced Metering Infrastructure (AMI) in power distribution systems \cite{Hidalgo2010}. Thus, system operators' access to residential, commercial, and industrial customer metering data has presented an opportunity for using data-driven techniques for system monitoring and control \cite{Mohassel2014}. While the AMI data history can be humongous in size, it does not necessarily provide full observability for distribution systems due to the limited number of smart meters compared to the huge size of the network and the common missing data problem \cite{Bhela}\cite{Miranda2012}. 

Pseudo-measurement generation techniques are used to improve the observability of distribution networks by performing data-driven power consumption estimation (in case of missing data, communication delays, and unobserved customers) \cite{Primadianto2017}. Also, weights are assigned to these estimated values to define the operator's confidence in the accuracy of pseudo-measurements in the state estimation process. Since the efficiency of distribution system control and management can be negatively affected by the inaccuracy of the generated pseudo-measurement samples,it is of critical importance to design data-driven load estimation methods capable of providing accurate pseudo-measurement samples to improve the quality of distribution system monitoring \cite{Alimardani2015}. 

Several papers have studied the problem of pseudo-measurement generation for distribution system monitoring and state estimation. The literature in this area can be roughly categorized into two groups based on the proposed solution approaches: 1) \textit{Statistical and probabilistic models:} The previous works in this category rely on statistical and probabilistic analysis of the available AMI data history for constructing pseudo-measurement generation methods. Empirical Gaussian distributions have been conventionally used for estimating the Probability Density Functions (PDF) of consumer load profiles and generating pseudo-measurements \cite{Arefi2015}. In \cite{Ghosh1997}, empirical customer consumption PDFs are constructed employing Beta and log-normal distributions, which show improved performance over single Gaussian approach. These PDFs are then used for generating estimated power consumption data samples. Gaussian Mixture Models (GMM) have also been shown to be an improvement over mere fitting of a single distribution function to the available data \cite{Singh2009, Singh2010}. In a more recent work, data clustering has been combined with GMM to improve the pseudo-measurement generation process \cite{Ghahrooei}. A weather-dependent empirical PDF construction scheme for distributed PV systems is proposed in \cite{Angioni2016-2}, as pseudo-measurement generator, which is shown to have superior performance over conventional statistical methods. Statistical load profile and power loss estimation have been used in \cite{Nguyen2015} and \cite{Chen2016}, respectively, to model the uncertainty of customer behavior and improve the observability of distribution networks. 2) \textit{Machine learning models:} Another group of researchers have adopted machine-learning-based methods for distribution system load estimation. In comparison with the first group, these methods are able to further improve the accuracy of pseudo-measurements by exploiting the available real-time data samples. A Probabilistic Neural Network (PNN) is proposed in \cite{Gerbec2005} for assigning load profiles to customers in distribution systems. In \cite{Manitsas2012}, an Artificial Neural Network (ANN) is used for generating pseudo-measurements using the real-time line power flow measurements. Missing data reconstruction using a neural network approach has also been employed in \cite{Miranda2012}. Using the concept of Parallel Distributed Processing networks (PDP) a load estimation mechanism has been developed in \cite{Wu2013} to design a robust state estimator for distribution systems. An adaptive Nonlinear Auto-Regressive eXogenous (NARX) model is proposed in \cite{Hayes2015} for load estimation in distribution networks. While these works provide invaluable insights into distribution system monitoring, they have certain shortcomings, including: failure to capture seasonal correlations in customer behavior, not addressing the big-data challenge for large AMI datasets, and ignoring the possibility of using Distribution System State Estimation (DSSE) data for improving machine learning performance.

In this paper, we propose a novel customer power consumption estimation process that can be used for pseudo-measurement generation for reconstructing unknown and missing smart meter data to improve the accuracy and precision of DSSE, while improving system observability. The proposed machine-learning-based method employs the concept of Relevance Vector Machine (RVM) to design sparse kernelized nonlinear regression models \cite{Tipping2001}. Moreover, unlike most regression models, RVM is capable of quantifying the uncertainty of pseudo-measurements by learning the variance of the estimated output. The variance learning process eliminates the need for relying on high-variance empirical distributions and is used to define weights for pseudo-measurements in the DSSE. Moreover, the inherent pruning mechanism of RVM introduces robustness against bad training data samples in the state estimation process. To alleviate the high cost of training, we propose a parallel computational framework using Multiple RVM (MRVM) units, each fitting a probabilistic model to a region of training set. The outcomes of these parallel training units are then recombined using a game-theoretic strategy to obtain final pseudo-measurement power consumption samples (along with their estimated variance). This game-theoretic framework is based on the concept of repeated games with vector payoffs \cite{Maschler2013} \cite{Bianchi2006}. It is observed that by employing this technique the pseudo-measurement generation accuracy can be significantly improved by exploiting the strong seasonal changes in customer behavior. The power consumption estimation model is then integrated with a Branch Current State Estimator (BCSE) module through a closed-loop information system to iteratively improve the pseudo-measurements using the additional information provided by the BCSE. The idea of using corrective closed-loop information system for DSSE has been employed in \cite{Hayes2015} and \cite{Wu2013}, as well. However, the novel approach in this paper is to train machine learning models using power-flow-based artificially-generated power consumption signals, which are expected to have high correlation with real customer consumption and improve the performance of pseudo-measurements. It will be shown that using the proposed approach, the performance of both pseudo-measurement generation and DSSE can be enhanced considerably compared to the open-loop case. The machine-learning-based estimation technique is tested on real data from a distribution feeder belonging to a utility company in the U.S. with smart meter measurements (power consumption and voltage measurement data). 

To summarize, the contributions of this paper are as follows: a novel computationally-efficient machine learning framework is proposed for enhancing the observability of distribution systems, and introducing robustness against bad training data. Seasonal changes in customer behavior are captured via a game-theoretic platform to improve the quality of pseudo-measurement generation. A closed-loop DSSE is developed for distribution systems, which highly improves the accuracy of state estimation by feeding the DSSE information back into the machine-learning-based power consumption estimation. The proposed framework is tested on real utility data for verification.

The rest of this paper is constructed as follows: in Section II, a description of the game-theoretic probabilistic learning framework for power consumption pseudo-measurement generation is presented. In Section III, the overall closed-loop DSSE module is described (and summarized in Section IV). The numerical results are analyzed in Section V. In Section VI, the conclusions of the paper are presented.  
\section{Proposed Pseudo-Measurement Generator}
\subsection{AMI Dataset Description and Pre-Processing}
The available AMI data history contains the hourly power consumption (kW) and voltage magnitude measurement of 3000 customers (with more than 40,000 data samples per customer) connected to 10 distribution feeders, which are located in the U.S. The dataset spans a time period of around five consecutive years (2013-2018). While a few industrial and large commercial loads are included in the dataset, the majority of customers are residential and small commercial loads.

The data was initially processed to remove grossly erroneous data samples (which were eventually replaced by pseudo-measurements). The bad data removal process was defined by the deviation of data samples from the mean of consumption signal for each customer. Hence, the samples that fall outside of $\pm 5$ deviation from the mean are removed, as having grossly erroneous values. The dataset was divided into two separate subsets for training ($80\%$ of the total data) and testing ($20\%$ of the total data). K-fold cross-validation was performed (over the training set) to choose certain model parameters (e.g., kernel bandwidth) \cite{Friedman2001}. A basic statistical analysis was performed on the dataset to identify variables with high correlation levels. As discussed in \cite{Luan2015}, the power consumption variable has a relatively high correlation level with voltage magnitude at the same bus and the neighboring nodes. This was also observed in the distribution system under study in this paper, specifically for larger customers, for which close to unit correlation values were recorded. Hence, available voltage magnitude measurement samples can be used as inputs in the regression models for estimating the consumption levels at different buses of the feeder. All these variables are normalized based on their maximum/minimum range of change. The objective of the pseudo-measurement generation process is to use the available noisy observations in real-time (i.e., voltage measurements, power measurements, time of day, etc.) to infer the unknown power consumption levels of unobserved customers (due to unavailability of meters, missing data, bad data, communication delays, and faulty measurements). To perform this task, regression models are trained using the system data history and employed to develop a mapping between the input and output data samples.

\subsection{Machine Learning Framework}\label{sec:ml}
\begin{figure}
      \centering
      \includegraphics[width=0.8\columnwidth]{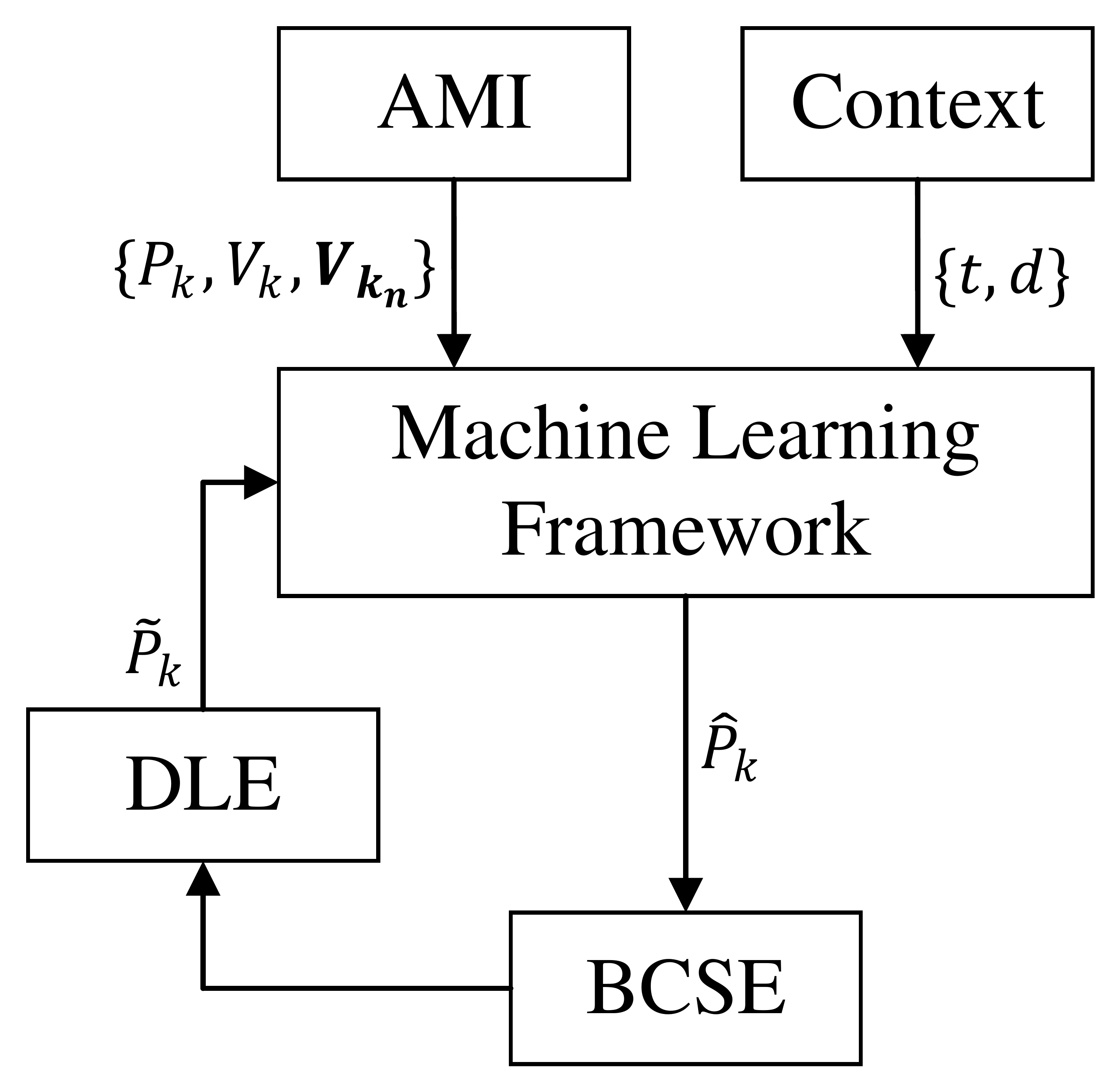}
\caption{Machine learning framework functionality.}
\label{fig:ML}
\end{figure}
The machine learning framework is a supervised approach, which maps the input data to the output target space (power consumption), as shown in Fig. \ref{fig:ML}. The input space consists of three types of variables: 1) AMI nodal voltage measurement data input with high correlation with customer power consumption (measured at bus $k$ ($V_k$) or neighboring buses ($V_{k_n}, k_n\in N_k$)), 2) context variables (time of day ($t$), and day of week $d$), and 3) the ``feedback'' power consumption signal generated by DSSE-based Load Estimation module (DLE), which also is highly correlated with the target power consumption (more details in Section \ref{Sec:SE}). Note that in this paper two distinct variables are defined to approximate the target power consumption space: $\hat{P}_k$, which defines the $k^{th}$ customer power pseudo-measurement variable (i.e., output of the machine learning framework), and $\tilde{P}_k$, which denotes the estimated customer power using the DLE module (i.e., DSSE feedback signal). Basically, after solving BCSE the estimated nodal voltages (or branch currents) are used to determine nodal power consumption levels. These estimations are used in a closed-loop mechanism to re-train the machine learning consumption estimation models. Hence, the role of the DLE module is to provide a link between the BCSE and machine learning framework. 

The RVM algorithm is premised on a kernelized regression model, which can be formulated as follows \cite{Tipping2001}:
\begin{equation}
\label{eq:regression}
\hat{P}_k = \sum_{i = 1}^{N} \omega_iK(\pmb{x}(k),\pmb{x_i}(k)) + \omega_0
\end{equation}
where, $\hat{P}_k$ represents the power consumption pseudo-measurement for the $k^{th}$ customer, $N$ denotes the total number of samples in the training set (i.e., number of previous observations), $\omega_i$ is the weight assigned to the $i^{th}$ input sample in the training set ($\pmb{x_i}$), and $K$ denotes the kernel function over the samples in the training set and the new input sample $\pmb{x}$ in the test set ($\pmb{x}(k) = \{V_k,\pmb{V_{k_n}},t,d,\tilde{P_k}\}$). In this paper, radial basis function kernel, which is a measure of similarity between the training samples and the new observations, is used to quantify $K(.,.)$:
\begin{equation}
\label{eq:kernel}
K(\pmb{x_i}(k),\pmb{x_j}(k)) = \exp\{-\frac{\norm{\pmb{x_i}(k)-\pmb{x_j}(k)}^2}{r^2}\}
\end{equation}
where, $r$ is a tunable parameter which defines the kernel bandwidth. The objective of the machine learning framework is twofold: 1) learn the parameters of the kernelized regression model ($\omega_i$'s), 2) quantify the uncertainty of estimation. This uncertainty is defined by the variance ($\sigma^2$) of the estimation error $\epsilon = P_k - \hat{P}_k$, where $P_k$ is the real power consumption of the $k^{th}$ user. RVM provides a computationally robust approach to achieve these goals. The learning mechanism employs a probabilistic view of the regression equation (\ref{eq:regression}), in which parameters $\pmb{\omega}= \{\omega_0,...,\omega_N\}$ are assumed to be normally-distributed independent random variables, with hyperparameters $\alpha_i$ defining their variance, as follows:
\begin{equation}
\label{eq:omegadist}
p\{\pmb{\omega}|(\alpha_0,...,\alpha_N)\} = \prod_{i = 0}^{N}\mathcal{N}(0,\alpha_i^{-1})
\end{equation}
where, $\mathcal{N}(a,b)$ denotes a normal distribution with mean $a$ and variance $b$. Note that using (\ref{eq:omegadist}), the $\alpha$ values can be used for eliminating irrelevant samples and pruning the training set. Accordingly, data samples for which the $\alpha$ levels converge to very large values can be removed safely from the training set, as their assigned weights get more concentrated around zero. The learning process is based on finding the most probable values for the set of hyperparameters $\{\alpha_0,...,\alpha_N\}$ and parameter $\sigma$ of the kernelized model to maximize the marginal likelihood function, which is formulated as follows:
\begin{equation}
\label{eq:marginlike}
(\pmb{\alpha^*},\sigma^*) = \argmaxA_{\pmb{\alpha},\sigma}p\{P_k|(\pmb{\alpha},\sigma)\}
\end{equation}

To achieve this, different recursive update rules have been obtained for these variables based on expectation-maximization process. The overall algorithm has the following steps for each bus, as discussed in \cite{Tipping2001}:
\begin{itemize}
\item \textbf{Step 1:} Initialize hyperparameters $\pmb{\alpha}$, and parameter $\sigma$
\item \textbf{Step 2:} Formulate the ``design matrix'', $\Phi$, and auxiliary matrix $A$ over the existing data samples in the training set $X = \{\pmb{x_1},...,\pmb{x_N}\}$:
\begin{equation}
\label{eq:design}
\mathbf{\Phi} = \left[
\begin{array}{cccc}
1 & K(\pmb{x_1},\pmb{x_1}) &  \cdots & K(\pmb{x_1},\pmb{x_N})\\
\vdots & \vdots & \ddots & \vdots\\
1 & K(\pmb{x_N},\pmb{x_1}) & \cdots & K(\pmb{x_N},\pmb{x_N})
\end{array}
\right]
\end{equation}
\begin{equation}
\label{eq:A}
A = \left[
\begin{array}{ccc}
\alpha_0 & & \\
& \ddots &\\
&&\alpha_N
\end{array}
\right]
\end{equation}
\item\textbf{Step 3:} Given the current values of $\pmb{\alpha}$ and $\sigma$, the parameters $\pmb{\omega}$ are estimated using a joint Gaussian distribution with covariance matrix $\pmb{\Sigma}$ and mean vector $\pmb{\mu}$, obtained as follows:
\begin{equation}
\label{eq:cov}
\pmb{\Sigma} = (\sigma^{-2}\Phi^\top\Phi + A)^{-1}
\end{equation}
\begin{equation}
\label{eq:mean}
\pmb{\mu} = \sigma^{-2}\pmb{\Sigma}\Phi^\top\pmb{P}_k
\end{equation}
\item\textbf{Step 4:} Update hyperparameters $\pmb{\alpha}$ and parameter $\sigma$ by equating the derivative of the objective function in (\ref{eq:marginlike}) to zero, as follows:
\begin{equation}
\label{eq:newalpha}
\alpha_i^{new}=\frac{1-\alpha_i\Sigma_{i,i}}{\mu_i^2}
\end{equation}
\begin{equation}
\label{eq:newsigma}
(\sigma^2)^{new} = \frac{||\pmb{P}_k - \Phi\mu||^2}{N-\sum_{i}(1-\alpha_i\Sigma_{i,i})}
\end{equation}
where, $\Sigma_{i,i}$ and $\mu_i$ denote the $(i,i)^{th}$ and $i^{th}$ elements of $\pmb{\Sigma}$ and $\pmb{\mu}$, respectively.
\item\textbf{Step 5:} Prune the training data set by removing samples that correspond to $\alpha_i \geq \alpha_{max}$, with $\alpha_{max}$ denoting a user-defined threshold. The columns and rows of $\Phi$ corresponding to the pruned data samples will also be removed.
\item\textbf{Step 6:} Go to Step 2, until convergence is achieved (i.e., changes in hyperparameters fall below a threshold.)
\end{itemize}

The objective of RVM is to learn a ``sparse'' model using the basic regression framework (\ref{eq:regression}) (with $\omega_i$'s and $\sigma$ as model parameters to be learned). The sparsity of the learning process is based on convergence of most of the model parameters ($\omega_i$) to near-zero values, which is also an automatic mechanism to avoid overfitting. To implement this mechanism, a pruning operation is performed at each iteration of the algorithm (Step 5) to eliminate the irrelevant data-points within the training set (only ``relevant'' samples are used for model-fitting.)

Following convergence, the estimated power consumption target variable at bus $k$ ($\hat{P}_k$) can be written as a conditional normal distribution (which is highly nonlinear in the input variables):
\begin{equation}
\label{eq:estimation}
p(\hat{P}_k|X) \sim \mathcal{N}(\pmb{\mu}^\top \pmb{\phi}(\pmb{x}(k)),\sigma^2+\pmb{\phi}(\pmb{x}(k))^\top \pmb{\Sigma\phi}(\pmb{x}(k)))
\end{equation}
where, $\pmb{x}(k)$ denotes the input variable from the test set. Also, $\pmb{\phi}$ is the basis function designed over the remaining training samples $\pmb{x_r} = \{\pmb{x_{r_1}},...,\pmb{x_{r_M}}\}$ where $\pmb{x_r}\subset X$, and is defined as follows:
\begin{equation}
\label{eq:basis}
\pmb{\phi}(\pmb{x}(k)) = \left[
\begin{array}{cccc}
1 & K(\pmb{x}(k),\pmb{x_{r_1}}) &  \cdots & K(\pmb{x}(k),\pmb{x_{r_M}})\\
\end{array}
\right]^\top
\end{equation}

As can be seen from (\ref{eq:estimation}), RVM is able to estimate both the target variable and its uncertainty (i.e., variance parameter, which represent factors such as noisy data and modeling errors.)
\subsection{Game-Theoretic Extension}\label{sec:game}
The computational complexity of RVM is normally proportional to $N^3$ (with $N$ denoting the number of training samples), which poses a considerable burden for AMI large datasets. In this paper, to reduce the high computational cost of learning, the training dataset is decomposed into multiple subsets and distributed among a population of RVMs that train models in parallel with each other. Hence, each RVM unit is trained based on a specific time interval of the input space. In this way, the computational load becomes proportional to $\frac{N^3}{M^2}$, with $M$ denoting the number of parallel RVM units. Hence, the computational complexity can be reduced by a factor of $1/M^2$ due to parallelization compared to the case where the whole dataset is used for training one RVM unit. The generated pseudo-measurement samples from the parallel RVM units are then recombined through weighted averaging (with weight value $w_{j,t}$ for the $j^{th}$ RVM unit at time $t$) to reach a final power consumption pseudo-measurement value. The objective is to find the optimal values of the weight values to maximize the pseudo-measurement accuracy. It was observed that to reach the best pseudo-measurement accuracy, the training set should be decomposed based on seasons of the year, which implies existence of strong seasonal changes in customers' behavior. Thus, four parallel RVM units (each corresponding to a season) are selected and trained over the training set. The recombination process has to be performed in a manner to preserve the precision of the estimation process. To perform this recombination task, the pseudo-measurement generation process is modeled as a repeated game with vector payoff \cite{Bianchi2006}. Based on this model, the game has two elements: 1) the ``nature'', which generates target time-series according to an unknown process (in our case, these time-series are the estimated customer consumption data generated by DLE), and 2) the ``estimator'' (referred to as the ``player''), which has the objective of inferring the behavior of nature and tries to maximize its long-term payoff by predicting the time-series generated by the nature. The estimator has access to multiple sources of ``advice'' (generated by RVM units) and needs to combine the received advice in a way to optimize its behavior in the game. Mathematically, the goal of the estimator is to minimize the \textit{Cumulative Regret}, which is defined with respect to the $j^{th}$ advisor ($j\in \{1,...,M\}$), $k^{th}$ customer, at time $m$, as follows:
\begin{equation}
\label{eq:CR}
R_{j,k}(m) = \sum_{t = 1}^{m} \{\ell(\hat{P}_k(t),P_k(t)) - \ell(f_{j,k}(t),P_k(t))\}
\end{equation}
where, $f_{j,k}(k)$ is the $j^{th}$ advisor (i.e., RVM unit) estimation of the target variable ($P_k(t)$). The function $\ell(.,.)$ defines the loss level due to mis-estimation, and is defined as $\ell(x,y) = |x-y|$ (which is convex in its first variable). Hence, the cumulative regret at a certain time point represents the player's loss for not following a specific advisor's estimations up to that point.  For ease of reference, the player's instantaneous regret level with respect to the $j^{th}$ advisor at time $t$ is defined as follows:
\begin{equation}
\label{eq:regret}
r_{j,k}(t) = \ell(\hat{P}_k(t),P_k(t)) - \ell(f_{j,k}(t),P_k(t))
\end{equation}
Hence, the instantaneous regret vector and the regret vector are defined as, $\pmb{r_k(t)} = (r_{1,k}(t),...,r_{M,k}(t))^\top$ and $\pmb{R_k(m)} = \sum_{t=1}^{m}\pmb{r_k(t)}$, respectively. While $\pmb{r_k(t)}$ represents a vectorized representation of instantaneous regret in the advisor space, $\pmb{R_k(m)}$ quantifies the summation of these instantaneous vectors up to a point in time.

The objective of the player is to assign optimal weight values to the advisors. Thus, the combination process for obtaining pseudo-measurements relies on weighted averaging of the received estimations from the RVM units, as follows:
\begin{equation}
\label{eq:weightedav}
\hat{P}_{k}(t) = \frac{\sum_{j=1}^{M}w_{j,k}(t-1)f_{j,k}(t)}{\sum_{j=1}^{M}w_{j,k}(t-1)}
\end{equation}

\begin{figure}
      \centering
      \includegraphics[width=1\columnwidth]{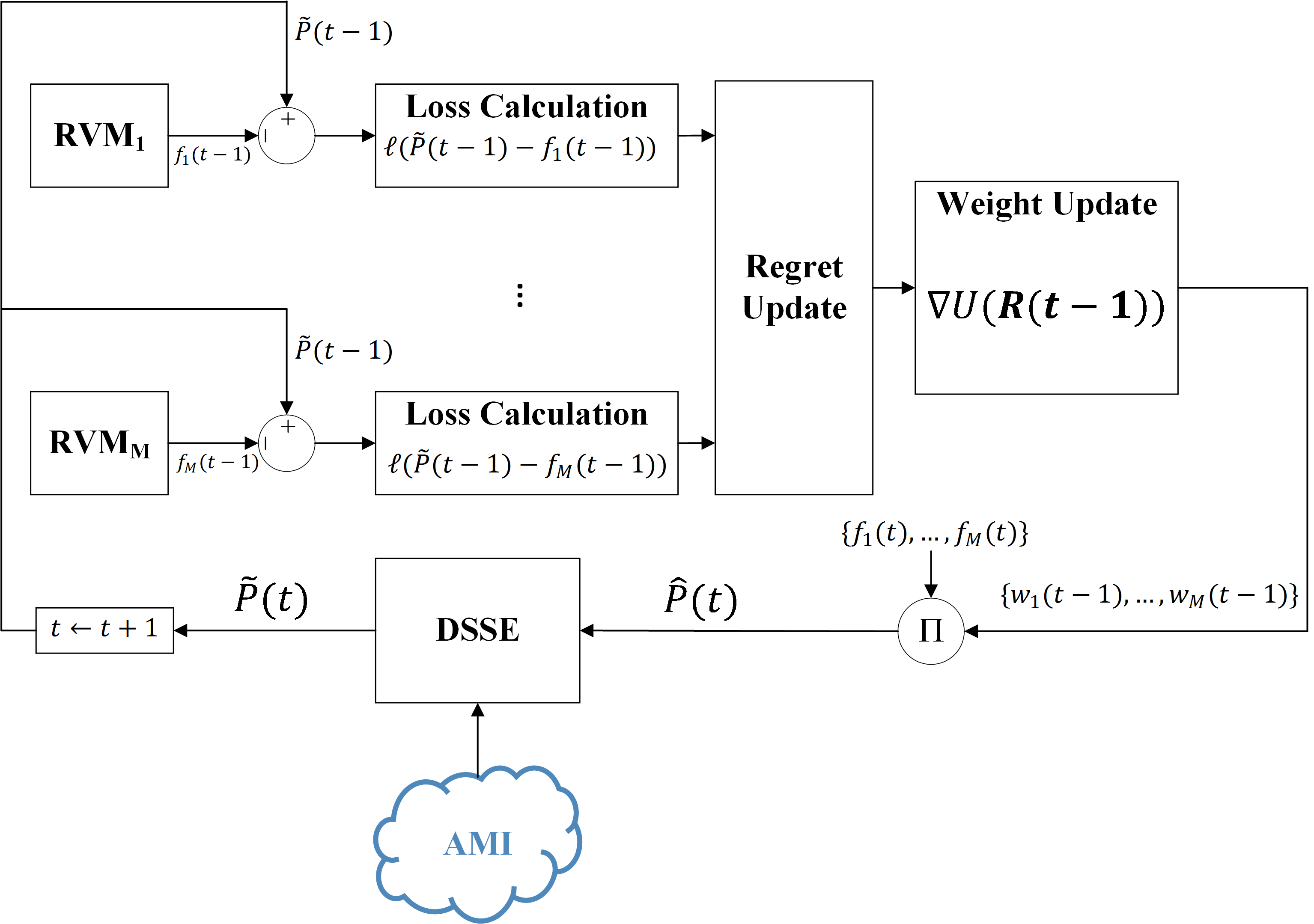}
\caption{Proposed structure of the game-theoretic learning process.}
\label{fig:game}
\end{figure}

The weight selection process is based on the choice of scalar non-negative, and twice-differentiable convex \textit{potential functions} over the regret vector, denoted by $U(\pmb{R_k(m)})$ \cite{Bianchi2006}. The goal of weight selection is to reduce the potential function value to limit the long term accumulated estimation regret. Basically, the potential function penalizes higher levels of regret. Hence, one choice of weight for adaptive correction of importance levels (weights) of RVM units is $w_{j,k}(t) = \nabla U(\pmb{R_{k}(t)})_j$ to improve the weights based on local gradient information of potential function. In this paper, an exponential potential function is chosen as follows:
\begin{equation}
\label{eq:potential}
U(\pmb{R_k(t)}) = \frac{1}{\eta_k(t)}\ln(\sum_{j=1}^Me^{\eta_k(t) R_{j,k}(t)})
\end{equation}
where, $\eta_k(t)$ is a tunable parameter (at time $t$). The choice of an exponential potential function leads to the following weight update mechanism:
\begin{equation}
\label{eq:expweight}
w_{j,k}(t-1) = \nabla U(\pmb{R_{k}(t-1)})_j = \frac{e^{\eta_k(t) R_{j,k}(t-1)}}{\sum_{j=1}^{M} e^{\eta_k(t) R_{j,k}(t-1)}}
\end{equation}
It can be proved that with the choice of $\eta_k(t) = \sqrt{\frac{8lnM}{t}}$ (and a normalized convex loss function) the following  upper-bound on the maximum regret level is achieved \cite{Bianchi2006}:
\begin{equation}
\label{eq:upper2}
\max_{j=1,...,M}R_{j,k} \leq 2\sqrt{\frac{k\cdot \ln M}{2}} + \sqrt{\frac{\ln M}{8}} 
\end{equation}
The overall game-theoretic platform is shown in Fig. \ref{fig:game}. As can be seen in this figure, the game-theoretic machine learning framework updates the importance weight factors online (in case the AMI samples or DLE outputs become available) or offline (using cross-validation). Also, the combined estimated pseudo-measurement variance for the $k^{th}$ customer ($\hat{\sigma}^2_k$) is calculated at time $t$ as follows:
\begin{equation}
\label{eq:varcomb}
\hat{\sigma}^2_k(t) = \frac{\sum_{j=1}^{M}w_{j,k}(t-1)^2\hat{\sigma}_{j,k}(t)^2}{(\sum_{j=1}^{M}w_{j,k}(t-1))^2}
\end{equation}
where, $\hat{\sigma}_{j,k}(t)^2$ is the estimated variance for the $j^{th}$ RVM unit at time $t$ obtained using (\ref{eq:estimation}).

\section{Closed-Loop DSSE Module}\label{Sec:SE}
\begin{figure*}
      \centering
      \includegraphics[width=2\columnwidth]{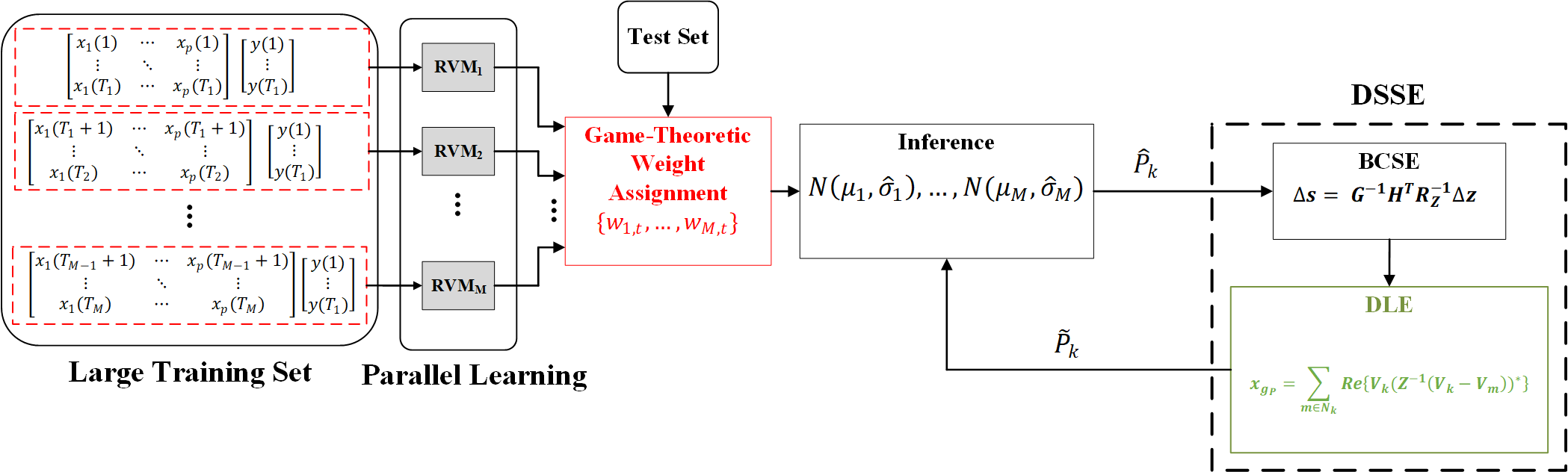}
\caption{Overall structure of the DSSE.}
\label{fig:overall}
\end{figure*}
The structure of the DSSE module is shown in Fig. \ref{fig:overall}. The module consists of two subsystems: BCSE and DLE.
\subsection{BCSE}\label{sec:bcse}
A BCSE algorithm is used for implementing the DSSE module \cite{Baran1995} \cite{Wang2004}. This algorithm is based on minimization of summation of weighted measurement residuals:
\begin{equation}
\label{eq:minJ}
\pmb{\hat{s}} = \argminA_{\pmb{s}} \sum_{i=1}^{N_z}\frac{1}{\sigma_i^2}(z_i-h_i(\pmb{s}))^2
\end{equation}
where, $z_i$'s represent the measurement and pseudo-measurements (with standard deviations $\sigma_i$ representing user's confidence, and total number of $N_z$), $\pmb{s}$ denotes the state vector, $h_i$ is the measurement function (which maps the state vector to the $i^{th}$ measurement/pseudo-measurement.) In this paper, the measurement samples are the active/reactive customer power consumption data, and PMU voltage measurement at the main substation. The state variables are the real/imaginary branch current values for each phase. Gauss-Newton method is used to iteratively update the state vector and achieve convergence. The update mechanism at step $q$ is as follows:
\begin{equation}
\label{eq:bcseupdate}
\pmb{s}_{q+1} = \pmb{s}_{q} + G^{-1}(\pmb{s}_{q})H^\top(\pmb{s}_{q})R_Z^{-1}(\pmb{z} - \pmb{h}(\pmb{s}_q))
\end{equation}
where, $G$ is the ``gain matrix'' defined as $G(\pmb{s}_{q}) = H^\top(\pmb{s}_{q})R_Z^{-1}(\pmb{s}_{q})H(\pmb{s}_{q})$, $H$ is the Jacobian matrix corresponding to the measurement function vector $\pmb{h}(\pmb{s}_q)$, and $R_Z = diag(\sigma_1^2,...,\sigma_{N_z}^2)$ is the measurement/pseudo-measurement uncertainty matrix.
\subsection{DLE}\label{sec:dle}
\begin{figure}
      \centering
      \includegraphics[width=0.95\columnwidth]{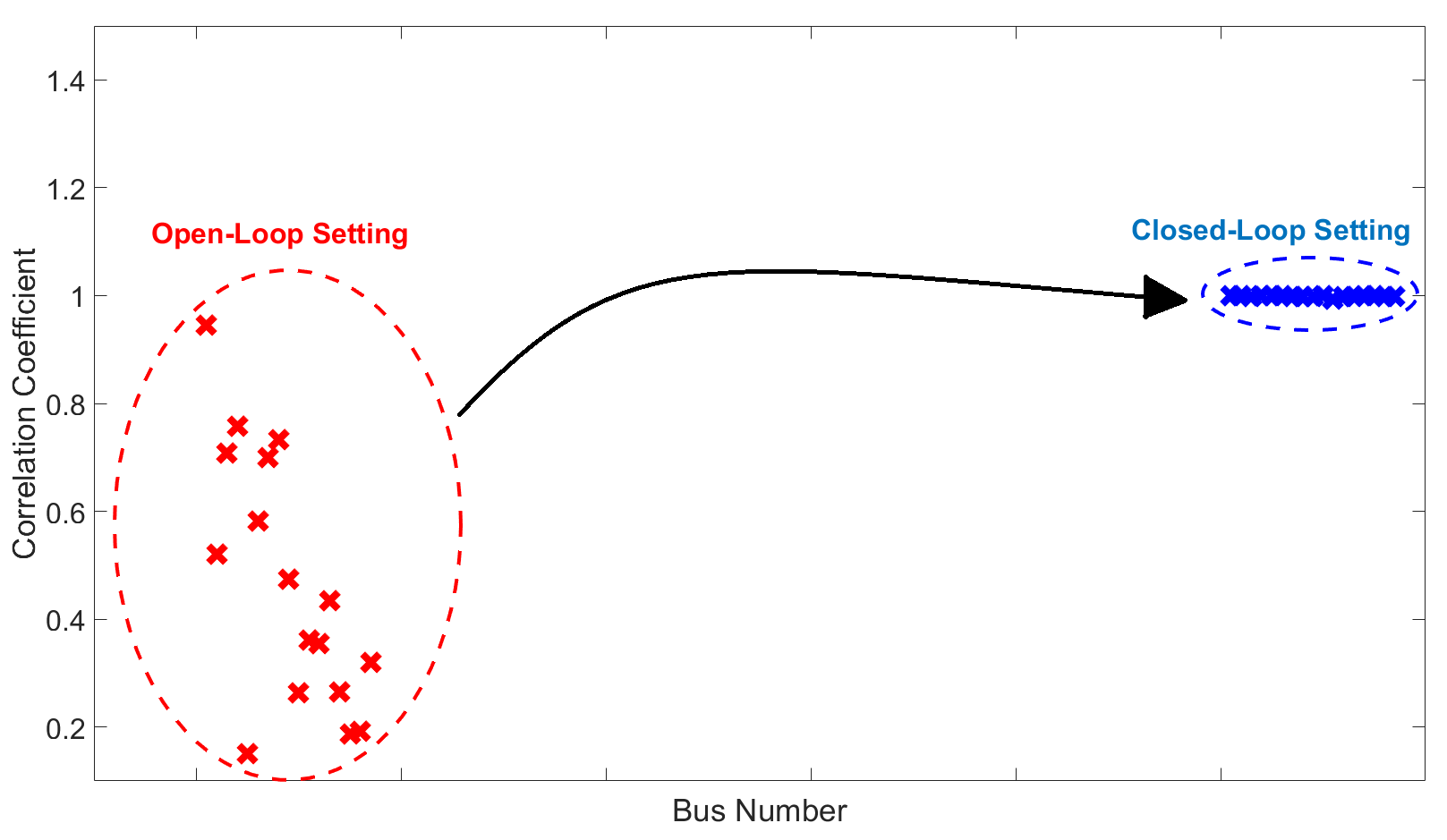}
\caption{Correlation between inputs/outputs of the machine learning framework, with respect to the state of the inner-loop.}
\label{fig:corr}
\end{figure}

After the convergence of the BCSE, the active power consumption is estimated at each node of the feeder using the estimated nodal voltage variables for each phase, employing power flow equations:
\begin{equation}
\label{eq:pf1}
\tilde{P}_k = \sum_{m\in N_k} \Re(\hat{V}_k(Z_{km}^{-1}(\hat{V}_k - \hat{V}_m)^*))
\end{equation}
where, $\hat{V}_k$ and $\hat{V}_m$ denote the BCSE-based three phase voltage phasor at bus $k$ and its neighboring nodes (included in the set $N_k$), and $Z_{km}$ defines the phase-based impedance matrix of the line connecting nodes $k$ and $m$. The estimated active power usage of each customer ($\tilde{P}_k$) is used to train and test the machine learning framework. The basic idea is that even under initial erroneous pseudo-measurement assignment, $\tilde{P}_k$ is highly correlated with the actual power usage information. The maximum correlation levels between the input/outputs of the machine learning framework at different nodes (for the primary distribution feeder) are shown in Fig. \ref{fig:corr}. As can be seen, the DLE output (obtained under open-loop state) has close-to-unity correlation with the actual power consumption. Hence, these artificially-constructed DLE signals can be exploited for training the machine learning framework to improve the accuracy of power consumption pseudo-measurements and state estimation algorithm in a closed-loop information system.   
\section{Overall Estimation Functionality}
In this section a summary of the different stages of the proposed state estimation framework is presented:
\begin{itemize}
\item \textbf{Stage I - Offline BCSE:} Perform BCSE on the measurement data history (consisting of real measurements and open-loop pseudo-measurement samples) to obtain estimated power consumption data.
\item \textbf{Stage II - Offline Training:} Augment the training set, using the DLE outcome of Stage I. Decompose the training set along seasonal time frames and train parallel RVM units (Section \ref{sec:ml}).
\item \textbf{Stage III - Weight Initialization:} Choose uniformly-random initial weights for the RVM units.
\item \textbf{Stage IV - Online Inference (time $T$):} Based on the available measurements and the DLE output (not available in the first iteration), and the weights assigned to RVM units update the value and weights of pseudo-measurements (Section \ref{sec:game}).
\item \textbf{Stage V - Online BCSE (time $T$):} Run the BCSE algorithm for $T$ based on the input measurements and pseudo-measurements until convergence is achieved (Section \ref{sec:bcse}).
\item \textbf{Stage VI - Online DLE (time $T$):} Update the power consumption information using the outcomes of DLE (Section \ref{sec:dle}).
\item \textbf{Stage VII - Loop Cycling (time $T$):} Go to Stage IV (with updates from DLE), until changes in pseudo-measurements for time $T$ fall below a threshold.
\item \textbf{Stage VIII - Weight Update:} $T\leftarrow T+1$, Update the weights assigned to the RVMs based on the latest available observations at time $T$ (Section \ref{sec:game}). Go to Stage IV.
\end{itemize}

The overall complexity of the proposed system monitoring can be approximated by $O(\frac{N^3}{M^2}+M+N_b^3f\epsilon^{-2})$, with $N_b$ denoting the number of distribution system nodes, $f$ is the number of iterations in the designed feedback loop, and $\epsilon$ is the threshold over gradient norm below which the BCSE is terminated. This complexity approximation is based on the computational complexities of three modules: multiple RVM learning ($O(\frac{N^3}{M^2})$) \cite{Tipping2001}, game-theoretic extension ($O(M)$) \cite{Bianchi2006}, and BCSE ($O(N_b^3f\epsilon^{-2})$).

The designed framework consists of numerical routines that need to have access to: 1) online AMI/SCADA/PMU data stream, and 2) AMI data history. In our case, the customer data history is available to utility partners directly or through hired third-party companies. The online data stream will be fed to the machine learning framework after resolving data formatting and structuring issues. Hence, protocols need to be designed to ensure the interoperability of interfaces. Other than that the proposed framework can be easily (and independently) implemented and integrated within the distribution automation systems with minimum modifications in the hardware (except maybe addition of parallel computational resources). The outcome of the framework is the state variables for the system operator.

\section{Numerical Results}
\begin{figure}
      \centering
      \includegraphics[width=0.9\columnwidth]{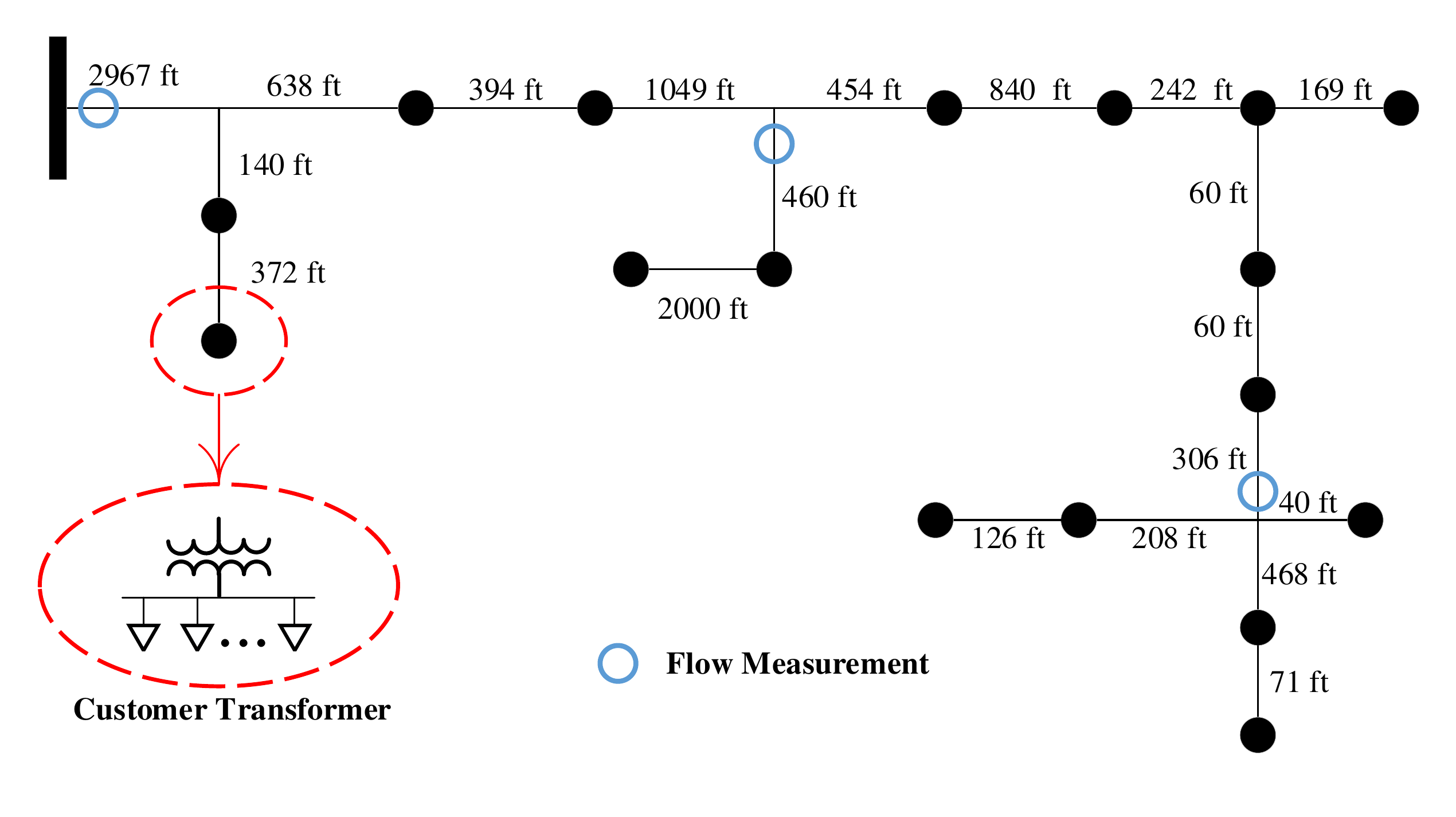}
\caption{The test system under study (220 customers).}
\label{fig:feeder}
\end{figure}

The proposed method is tested on a sample feeder from the available utility dataset (described in Section II) with 220 customers. The feeder is shown in Fig. \ref{fig:feeder}. The test feeder has three power flow measurement units and has around 35\% smart meter penetration. The accuracy of measurement units is assumed to be $\pm1\%$. Note that in this paper BCSE has been performed on the primary distribution network, since the secondary network topology data is not available. Hence, the power loss on the secondary network is assumed to be negligible and the consumption levels for the customers connected to the same transformer are simply added together at different times. If the topology data for the secondary distribution networks is acquired, the estimation process over these networks can be integrated into the state estimation algorithm with minimal or no changes to the proposed machine learning framework. Another way to look at this is to add a topology discovery step to the proposed machine learning framework to obtain a fully accurate BCSE module. The performance of the monitoring system is analyzed in both open- and closed-loop states. Also, the machine learning framework's robustness against bad data has been compared to conventional methods, such as ANN, linear regression, and Gaussian Maximum Likelihood Estimation (MLE). 
\subsection{Pseudo-Measurement Generation Performance}
\begin{figure}
      \centering
      \includegraphics[width=0.9\columnwidth]{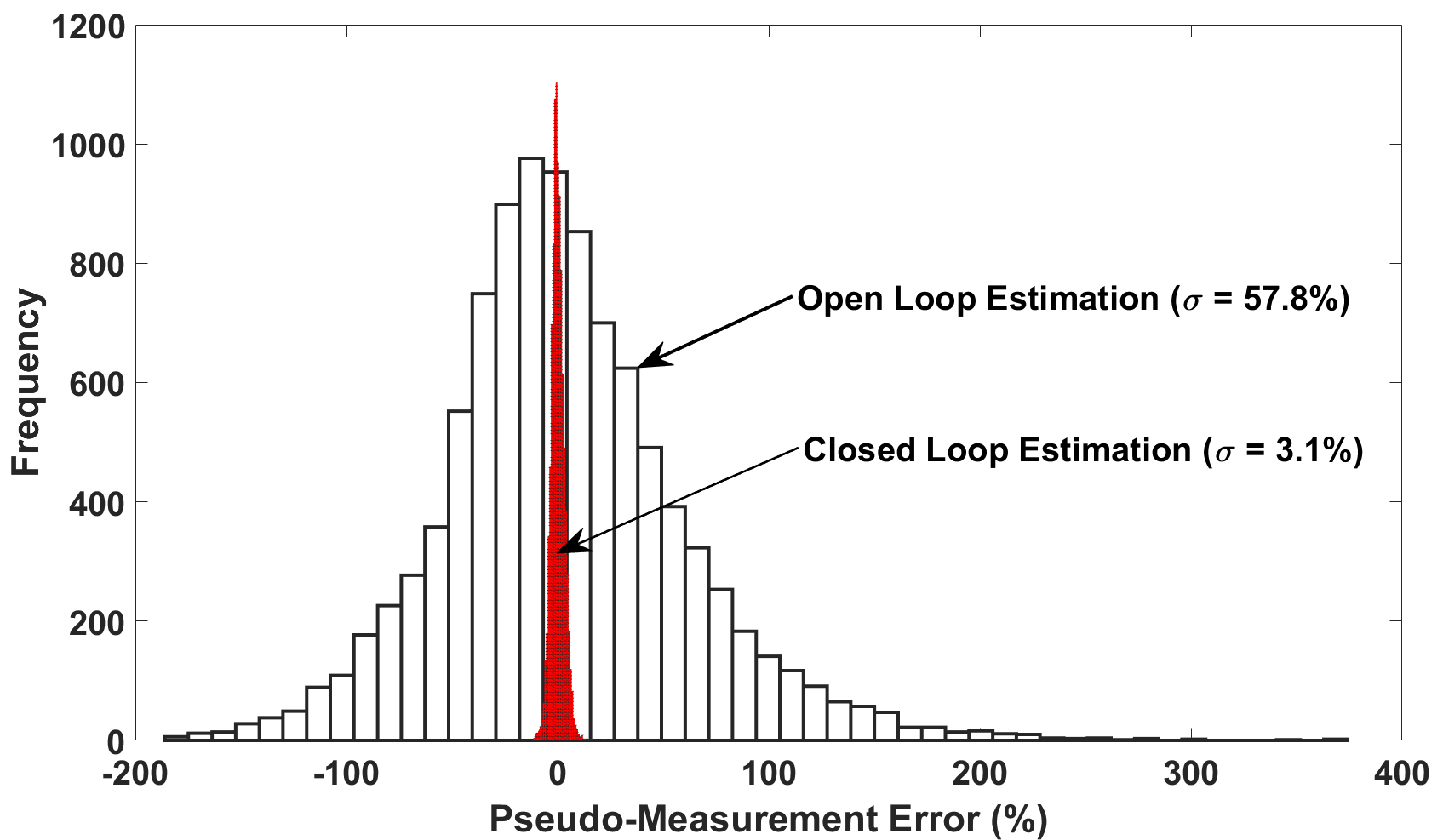}
\caption{The pseudo-measurement generation error histograms.}
\label{fig:hist}
\end{figure}

The machine learning framework was tested using the AMI data history. The histogram for customer consumption pseudo-measurement error is shown in Fig. \ref{fig:hist} for both open- and closed-loop situations. As can be seen in this figure, by closing the inner-loop (i.e., using DLE data) the pseudo-measurement \textit{precision} (defined as the inverse of the error distribution variance) has been improved by a considerable margin of 347.6\%. The Mean Absolute Percentage Error (MAPE) has also been reduced from 31.74\% to 1.94\% by employing model training using the signals generated by the DLE module in the inner-loop. The actual customer consumption is used as the ground truth for performance evaluation. 

\begin{figure}
      \centering
      \includegraphics[width=0.9\columnwidth]{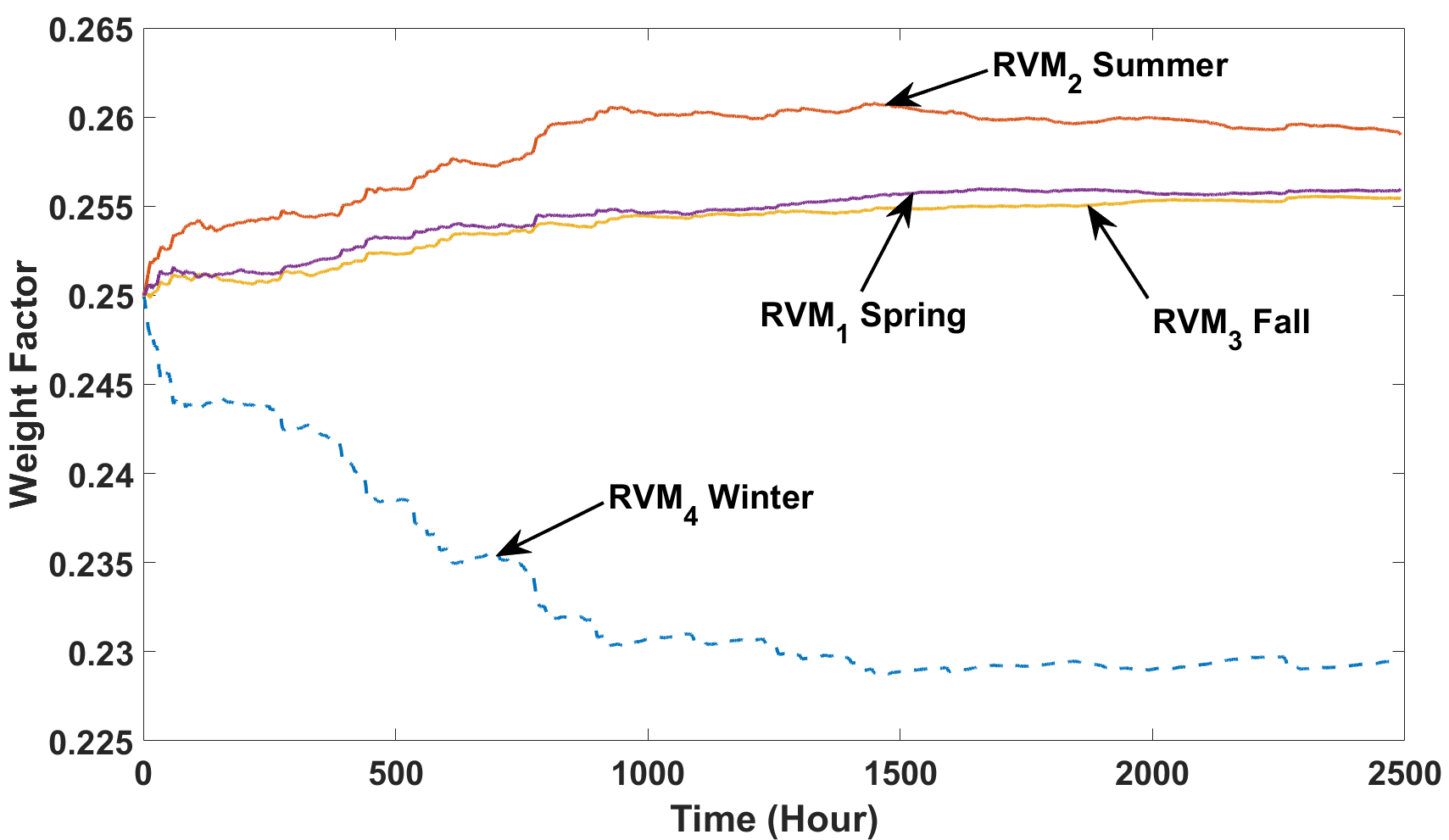}
\caption{Game-theoretic weight assignment (outer-loop).}
\label{fig:weights}
\end{figure}

The behavior of the outer-loop is captured by studying the changes in the game-theoretic weight assignment module. The weights assigned to the parallel RVM units (averaged over all customers), corresponding to different seasons of the year in the training set, are shown in Fig. \ref{fig:weights}. Given that the test set is selected to be the summer of 2017, higher weights are assigned to the regions of training set with similar patterns (summer and spring of 2014-2016). A critical aspect of the estimation process is that the game-theoretic aggregation of the RVM units outperforms each of the individual units in the long run \textit{on average}. The long run average MAPE for the aggregate estimator is 1.94\%, while this index increases to 2.18\%, 2.97\%, 3.15\%, and 5.37\% for the available individual RVM units, implying the advantage of the proposed signal combination method in terms of accuracy. Hence, parallelization not only reduces training computational complexity but also leads to more accurate pseudo-measurement samples. 

\begin{figure}
      \centering
      \includegraphics[width=0.9\columnwidth]{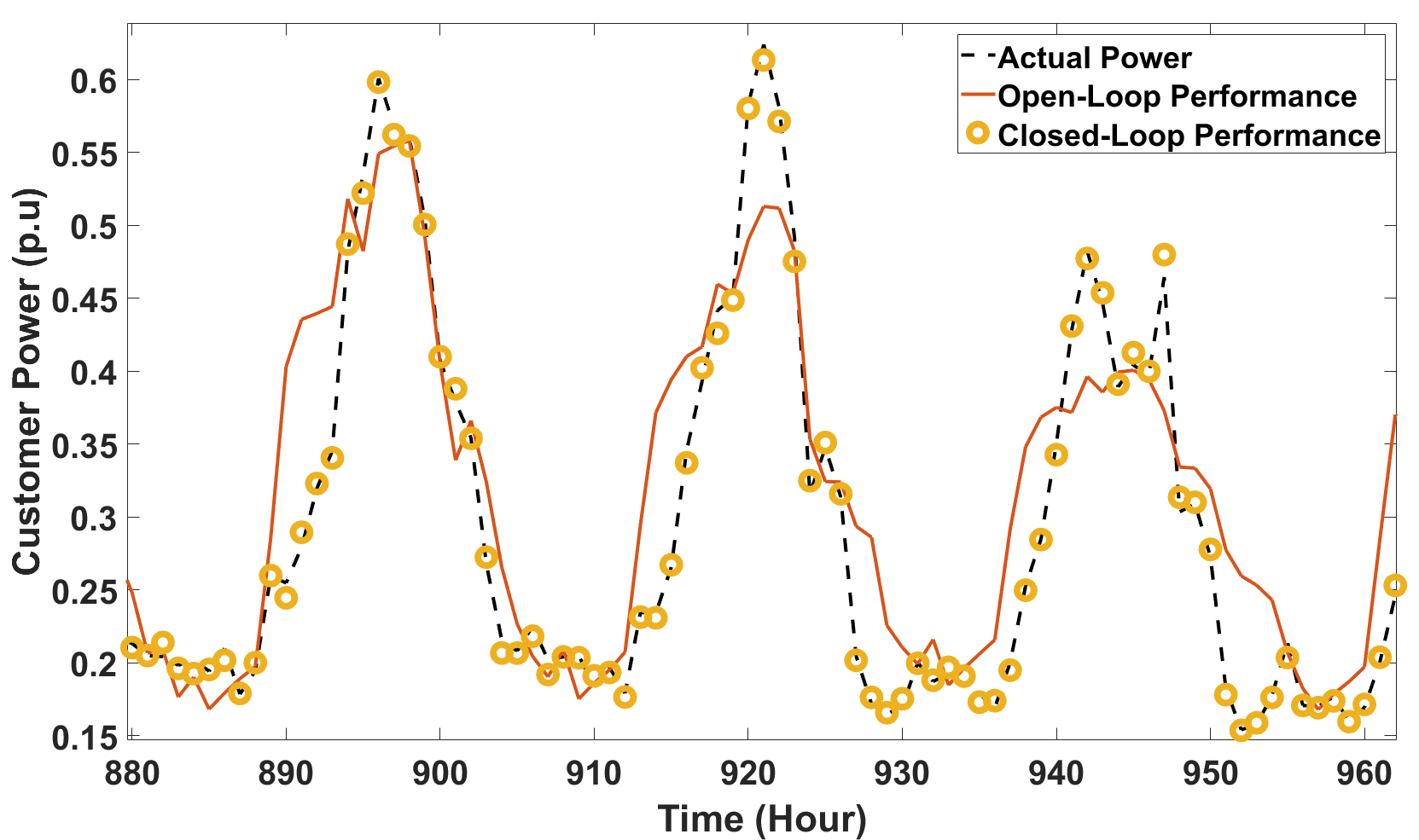}
\caption{Pseudo-measurement accuracy demonstration.}
\label{fig:pseudo}
\end{figure}

The performance of the pseudo-measurement generation module for the two cases of open and closed inner-loop states are shown in Fig. \ref{fig:pseudo}. As can be seen in this figure, after closing the inner-loop near-perfect fit to the underlying data can be achieved, which demonstrates the effectiveness of the proposed machine learning framework in closed-loop setting.

\begin{figure}
      \centering
      \includegraphics[width=0.9\columnwidth]{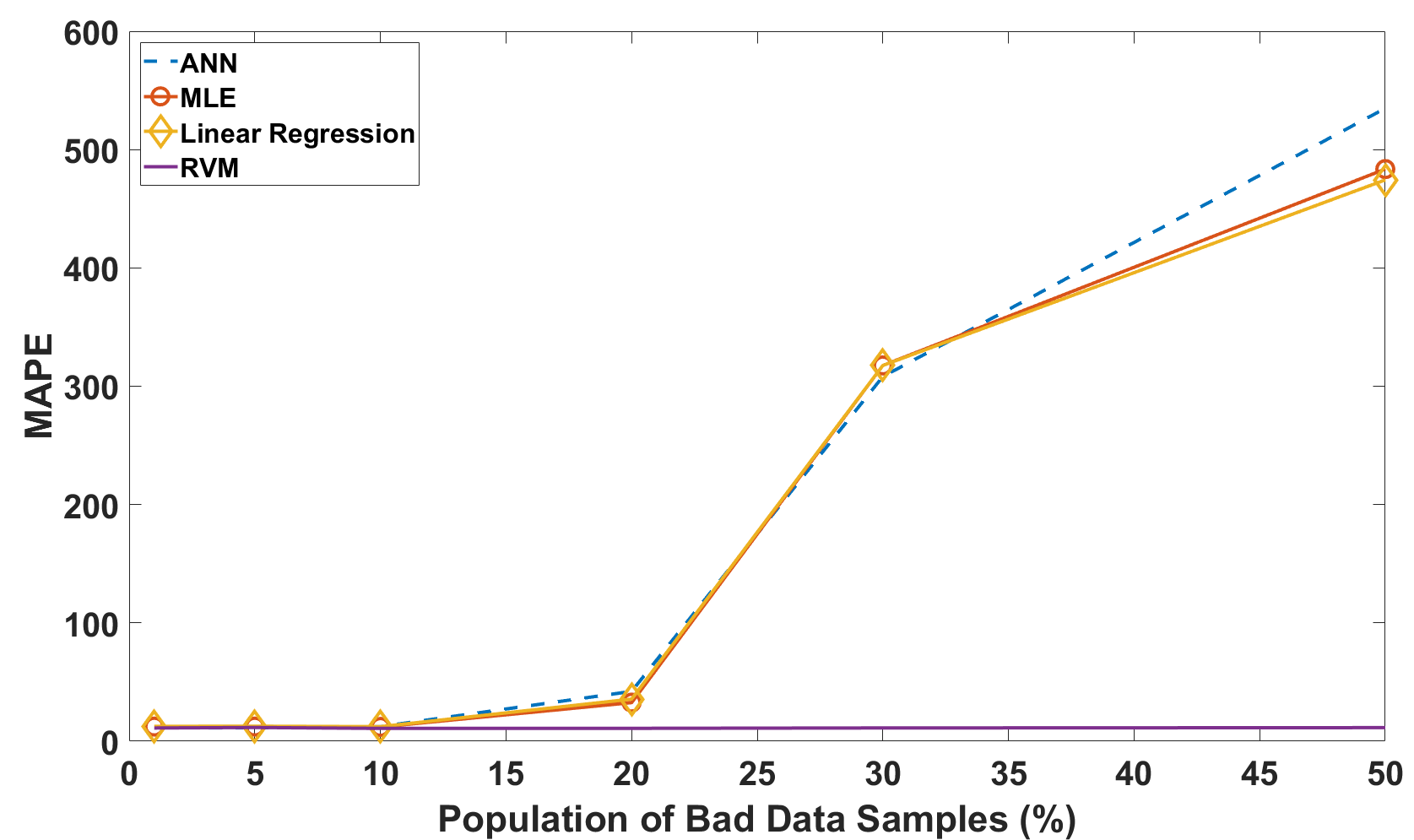}
\caption{Performance of the machine learning frameworks against bad data.}
\label{fig:bad}
\end{figure}

The robustness of the proposed machine learning model is tested by injecting artificially generated bad data to the training set. The pseudo-measurement generation MAPE is shown as a function of the bad data sample population for different methods in Fig. \ref{fig:bad}. To add the error to the training data two steps were taken: 1) $N$ data points were randomly selected from the training set. 2) Noise values generated by Gaussian distributions were added to each selected data point. The Gaussian distributions have zero means and standard deviations equal to 50\% of the magnitude of the corresponding selected data sample. After distorting the $N$ training data samples the machine learning models are trained and tested. This process is repeated several times for each $N$ value. Then $N$ is modified (decreased or increase). As is seen in this figure, an increase in the population of bad data samples leads to a drastic decline in the performance of ANN, MLE, and linear regression. However, the performance of the proposed MRVM method remains highly stable for a wide range of bad data sample population size. The reason for this stability is the ability of the RVM algorithm to prune the training dataset and eliminate ``irrelevant'' data samples that do not contribute positively to the marginal likelihood function. In other words, RVM has a natural mechanism for bad data detection and elimination, which is highly beneficial when dealing with real data. 

\subsection{State Estimation Performance}
\begin{figure}
\centering
\subfloat[Branch current real component error\label{sfig:real}]{
\includegraphics[width=0.45\textwidth]{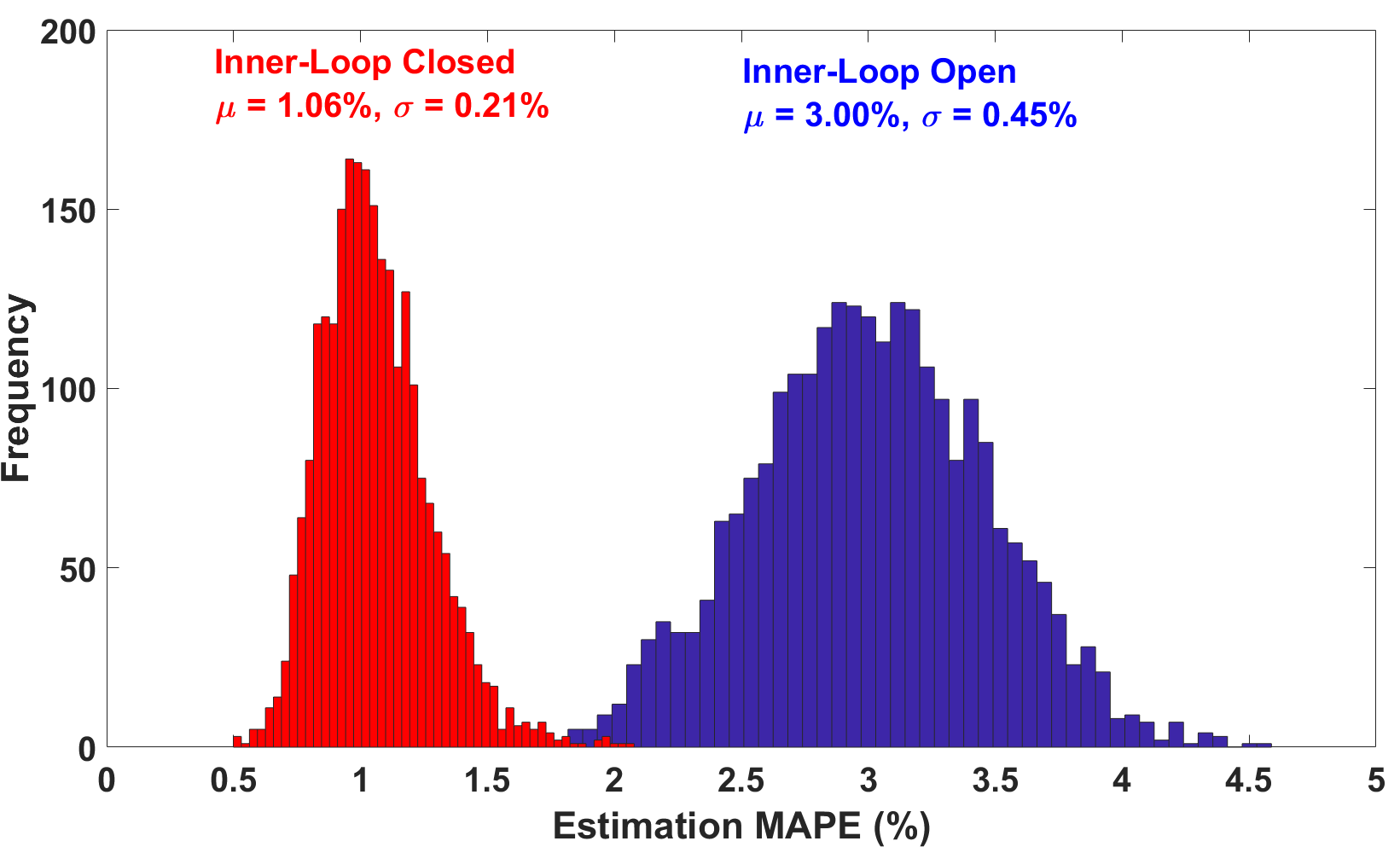}
}
\hfill
\subfloat[Branch current imaginary component error\label{sfig:img}]{
\includegraphics[width=0.45\textwidth]{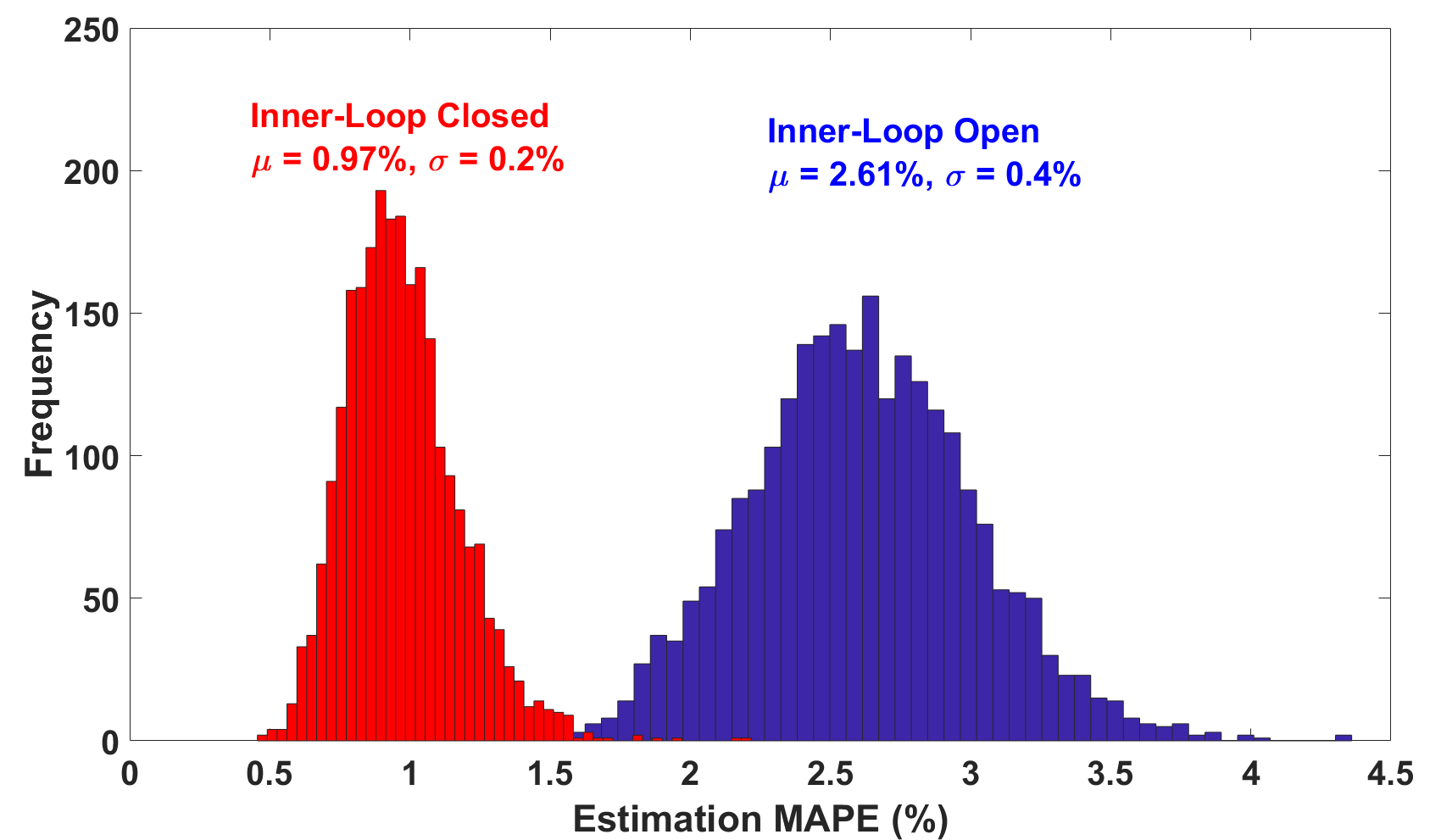}
}
\caption{BCSE performance in estimating state variables in open- and closed-inner-loop conditions.}
\label{fig:Ihist}
\end{figure}

The state estimation performance (in terms of MAPE) is shown in Fig. \ref{fig:Ihist} for both open- and close-loop conditions for real and imaginary branch current components. As is demonstrated in these figures, using the closed-loop DSSE module improves both the accuracy and precision (i.e., mean and variance) of the BCSE.   

\begin{figure}
      \centering
      \includegraphics[width=1\columnwidth]{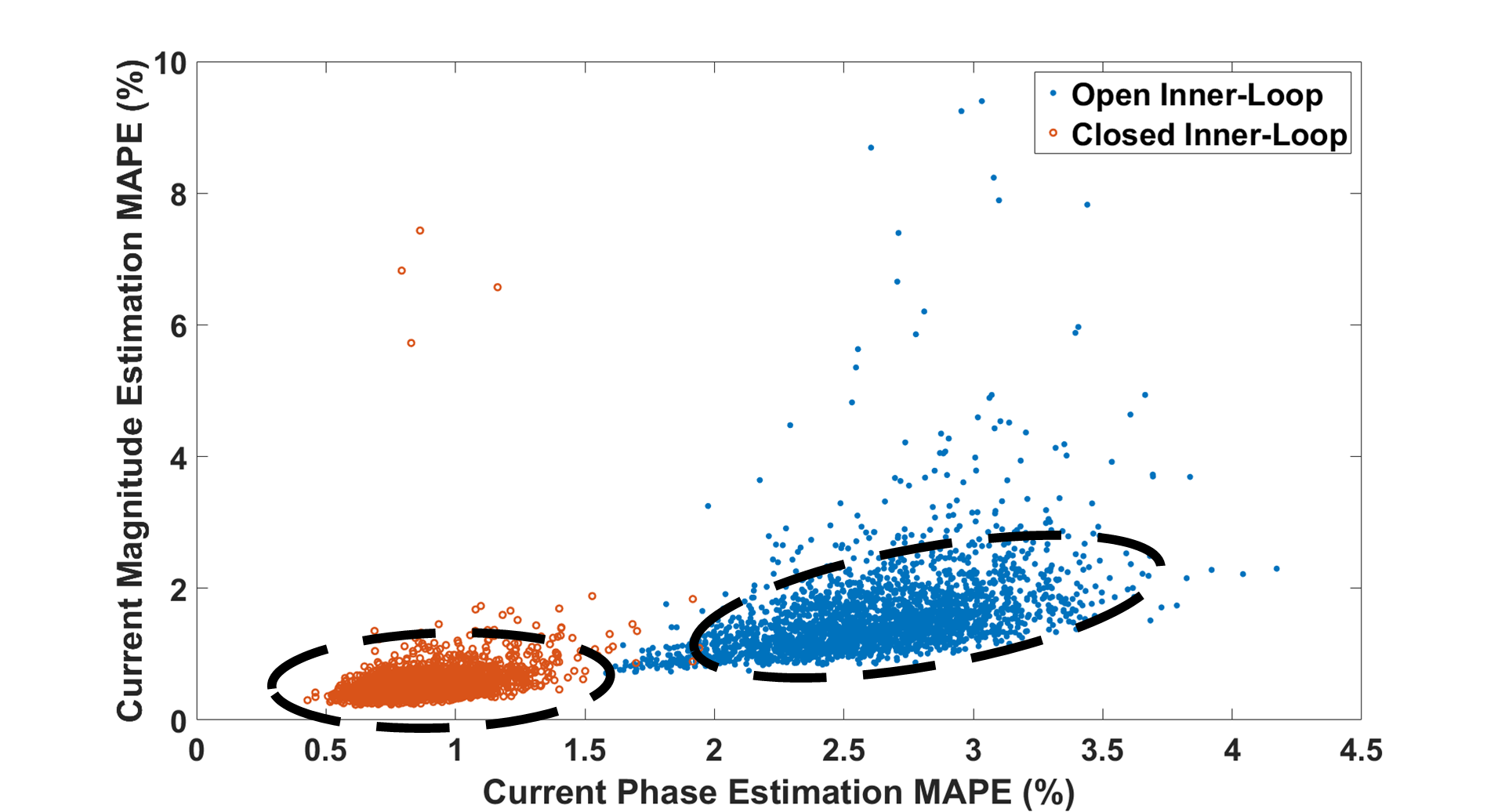}
\caption{State estimation accuracy demonstration (open-loop and closed-loop comparison).}
\label{fig:scatter}
\end{figure}

The distribution of current magnitude and phase estimation error is shown under open- and closed-inner loop conditions in Fig. \ref{fig:scatter} using scatter plots. In this figure, the improvements in DSSE can be observed, where a shift in the regions with high concentration of error data is observed (from (1.57\%,2.61\%) to (0.54\%,0.87\%)). We have also observed that the performance of state estimation depends on the location and number of measurement units distributed across the system. However, in all cases the proposed closed-loop machine learning framework leads to improvements compared to the open-loop setting for any number of measurement units.

\begin{figure}
      \centering
      \includegraphics[width=0.9\columnwidth]{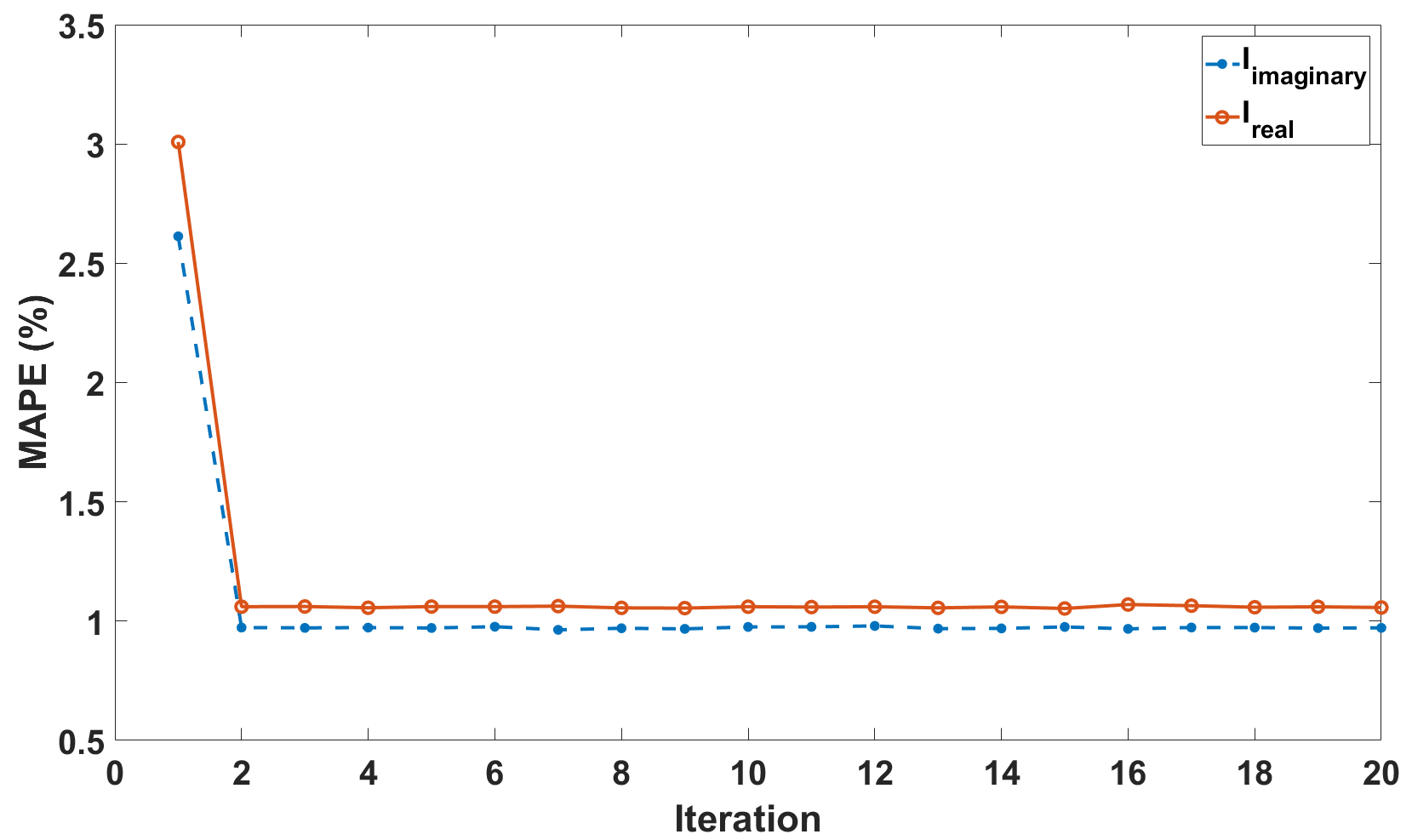}
\caption{Convergence of the proposed DSSE module.}
\label{fig:convergence}
\end{figure}

The convergence of the proposed DSSE model is shown in Fig. \ref{fig:convergence}, where the estimation MAPE is demonstrated as function of iterations, with each iteration representing a cycle in the inner-loop. Note that the estimation error is calculated as an average over all branches in the feeder. As is seen in the figure, the proposed method reaches steady-state after a single iteration, which implies fast convergence and suitability for real-time applications.

The proposed framework requires an average 10.1 seconds per transformer per year of training data to generate solutions for each hour, as tested on a Intel(R) Xeon(R) CPU E3-1240 V6 @ 3.7 GHz hardware. Hence, given that the processing time is almost 357 times faster than the actual system time flow, the proposed method is well capable of real-time monitoring of distribution system states. The total training time using the data collected over 3 years, is 484.2 seconds. 

\section{Conclusions}
In this paper, we have presented a computationally-efficient machine learning method for accurate pseudo-measurement generation to improve the quality of DSSE against unknown, missing, and bad data. The proposed approach is based on parallel training of multiple machine learning units and is shown to be highly robust against bad data samples in the training set. Employing the proposed technique we are able to exploit the seasonal patterns in customers' behavior to improve the accuracy of pseudo-measurement generation. A nested closed-loop DSSE module is developed to improve the accuracy and precision of the state estimation process by enabling interaction between the learning framework and the DSSE. The proposed method is successfully tested on a utility feeder with real smart meter data.

\ifCLASSOPTIONcaptionsoff
  \newpage
\fi



\bibliographystyle{IEEEtran}
\bibliography{IEEEabrv,./bibtex/bib/main}
\end{document}